\documentclass[aps,prd,showpacs,nofootinbib,floats,floatfix,preprintnumbers,groupedaddress,twocolumn]{revtex4}

\usepackage{bm}
\usepackage{latexsym}
\usepackage{dcolumn}
\usepackage{amsmath,amsfonts,amssymb}
\usepackage{graphicx,epsfig}
\usepackage{amsmath}
\usepackage{fancyhdr}
\usepackage{hyperref}
\usepackage{graphicx,epstopdf}
\usepackage{physics}

\hypersetup{
	colorlinks   = true, 
	urlcolor     = blue, 
	linkcolor    = blue, 
	citecolor   = red 
}

\begin{document}
\title{Are non-vacuum states much relevant for retrieving shock wave memory of spacetime?}
\author{Bibhas Ranjan Majhi\footnote {\color{blue} bibhas.majhi@iitg.ac.in}}

\affiliation{Department of Physics, Indian Institute of Technology Guwahati, Guwahati 781039, Assam, India}

\date{\today}

\begin{abstract}
Shock wave gives back reaction to spacetime and the information is stored in the memory of background. Quantum memory effect of localised shock wave on Minkowski metric  is investigated here. We find that the Wightman function for massless scalar field, on both sides of the wave location, turns out to be the same for usual Minkowski spacetime.  Therefore the observables obtained from this, for any kind of observer, do not show the signature of the shock. Moreover, the vacuum states of field on both sides are equivalent. On the contrary, the correlator for the non-vacuum state does memorise the classical shock wave and hence the effect of it is visible in the observables, even for any frame. We argue that rather than vacuum state, the non-vacuum ones are relevant to retrieve the quantum information of classical memory.
\end{abstract}


\maketitle

{\it Introduction and motivation} --
Contemporary works of Bekenstein \cite{Bekenstein:1973ur} and Hawking \cite{Hawking:1974rv} reveal that black holes are indeed thermodynamical objects. In identifying  the temperature of the horizon, Hawking effect \cite{Hawking:1974rv} plays a pivotal role. Although these have been verified again and again, a very fundamental issue related to the final stage of the black hole evaporation is still unsolved. After complete evaporation of the black hole, it is argued that the information about the collapsed mass should be encoded in the radiation, giving rise to a mixed state. But at this stage since ``in'' state is completely lost, the entanglement between the ``out'' and ``in'' states is no longer there. Then the appearance of this thermality at the final stage of evaporation is not properly justified. There are various attempts to address this unitarity problem, known as the {\it information paradox}. Existence of remnant \cite{Adler:2001vs,Banerjee:2010sd,Paul:2016xvb,Majhi:2013koa}, modification to quantum mechanics \cite{Modak:2014vya}, introducing fuzzball concept \cite{Mathur:2008kg} are among some of them.

A general belief is that the complete quantum analysis can give the answer to this question. In the absence of such ultimate theory, the calculations are so far semi-classical by nature. Even within this approximation, in most cases the back reaction of the collapsing matter has not been included. However following the works of Bondi-Burg-Metzner-Sachs (BMS) \cite{Bondi:1962px,Sachs:1962wk,Sachs:1962zza} (also see \cite{Newman:1966ub}), it has been recently observed that the asymptotic symmetry of the spacetime can lead to supertranslation charge \cite{Hawking:2016msc,Hawking:2016sgy}, known as `soft' hair. So at the classical level these soft hairs are observable and moreover the spacetime metric is being modified. Interestingly, this can be attributed as a result of propagation of shock wave in the spacetime. Then the modified spacetime can be visualised as the modified metric due to the back reaction of the propagating matter shell \cite{Hawking:2016sgy}. 

Hawking et. al \cite{Hawking:2016msc,Hawking:2016sgy,Strominger:2017aeh} argued that these soft hair must have quantum imprint, known as {\it quantum memory} and thereby may play a big role in solving the puzzle of information paradox. This led several investigations, starting from asymptotic symmetries of spacetimes \cite{Donnay:2015abr,Eling:2016qvx,Akhmedov:2017ftb,Maitra:2018saa} to its consequences \cite{Zhang:2017geq,Gomez:2017ioy,Kolekar:2017tge,Donnay:2018ckb,Chu:2018tzu,Javadinazhed:2018mle,Maitra:2019eix}. In this regard, the simplest toy model to study is the uniformly accelerated frame in the Minkowski spacetime. Although, this does not represent black hole scenario, but {\it equivalence principle} indicates that several features of gravity can be extracted from this simplified mathematical setup. Unruh effect \cite{Unruh:1976db} is the most well known phenomenon which mimics Hawking effect. In this spirit, it is important to investigate how classical soft hair for Minkowski spacetime can be memorised in quantum observables. 
The model we shall concentrate here is due to Dray and 't Hooft \cite{Dray:1984ha} -- shock wave Minkowski background (SMB). Investigation so far was confined within evaluation of number operator in SMB vacuum, as measured by the uniformly accelerated observer.  It shows that observer does not see any quantum imprint of the shock wave in the emission spectrum \cite{Compere:2019rof}. Interestingly the same has been seen in the computation of Hawking temperature for the soft hair induced Schwarzschild black hole as well \cite{Hawking:2016sgy,Javadinazhed:2018mle,Lin:2020gva}. Therefore this negative result has nothing to do with the curvature of spacetime.  Following an earlier work \cite{Chu:2018tzu} on Vaidya black hole, very recently it has been argued that the dynamical spacetime can have quantum memory \cite{Chiang:2020lem}. 

The common features of all these investigations is that only the vacuum state of the field has been studied. In this semi-classical attempt, although back reaction due to wave is being considered in the spacetime, the quantum fluctuation of the shock wave has not been considered. We may understand this as follows. Suppose the shock wave is made of scalar fields, described by operator $\phi(x) = \phi_0+\delta\phi$. The first part generates the classical shock wave and so it modifies the spacetime. Whereas other one represents small fluctuation which is not classically observed. Then this may always keep the field states as non-vacuum. Alternatively, one may think the shock wave is due to some excited matter.  A similar idea was introduced originally in \cite{Lochan:2015oba,Lochan:2016nbs} to extract the information about the ``in'' state in which back reaction was not included.      
In this situation the non-vacuum states of the fields might be the probable state to capture the memory effect at the quantum level. A small discussion has been done in \cite{Compere:2019rof} for one particle state to calculate the spectrum, observed by accelerated frame, but without any prior motivation.

So far the existing analysis is very restrictive in a sense that this is confined only to obtention of Bogoliubov coefficients. Moreover, such analysis has always done with respect to the accelerated frame. So the situation of the various observables and observers are not known to us. Given this, we are here to find a common quantum quantity from which not only various observables can be computed in SMB vacuum, but also the situation with respect to other observers can be investigated. One obvious such quantity is the two point correlation function for the quantum fields. This will not only help to understand the cause of non-existence of quantum memory in emission spectrum, but also will tell us whether there is any possible observable which encodes such memory.

In this letter, we first investigate the SMB vacuum for massless scalar fields on both sides of shock wave position. The positive Wightman function is found to be same as that for metric in absence of wave. This clarifies that the SMB vacuum does not attribute the quantum signature of the soft hair with respect to any observer in any of the observable computed from this. In addition we compared the vacua in both sides of the wave location. It is observed that they are equivalent. Finally, the non-vacuum state is being studied. We again find the two point function for one particle state. The features of the one particle state on both sides are found to be different and so they are inequivalent. Moreover, this state on the future of the wave captures the classical memory even with respect to the static observer. Therefore we argue that {\it the non-vacuum states of fields, in retrieving the quantum information of classical memory, is much efficient than the vacuum one}.

{\it Shock wave metric and massless scalar modes} --
The shock wave metric for Minkowski spacetime, following Dray and 't Hooft \cite{Dray:1984ha}, is given by
\begin{equation}
ds^2 = -dUdV+f({\bf{x}}_{\perp})\delta(U-U_0)dU^2 + d{\bf{x}}_{\perp}^2~,
\label{2.01}
\end{equation}
where $U=T-X$, $V=T+X$ are null coordinates and ${\bf x}_{\perp}$ are the transverse coordinates. This metric corresponds to the localised shock wave at $U=U_0$ which is propagating along positive $X$ direction and is represented by the stress-tensor $T_{UU}=\delta(U-U_0)T({\bf x}_{\perp})$. Here $T$ is a function of only transverse coordinates and satisfies $\nabla_{\perp}^2 f({\bf x}_{\perp}) = -16\pi G T({\bf x}_{\perp})$.

The massless scalar field mode solutions propagating on the metric (\ref{2.01}) are evaluated as \cite{Klimcik:1988az,Compere:2019rof}
\begin{equation}
g_{k_-, {\bf k}_{\perp}} = N_ke^{-ik_- V}e^{-ik_+(U-U_0)+i{\bf k}_{\perp}\cdot{\bf x}_{\perp}},~\textrm{for}~ U<U_0~;
\label{2.02}
\end{equation}
and
\begin{eqnarray}
&&f_{k_-, {\bf k}_{\perp}} = N_ke^{-ik_-V}\int \frac{d^2{\bf x'}_{\perp}}{(2\pi)^2} e^{i{\bf k}_{\perp}\cdot {\bf x'}_{\perp}+ik_-f({\bf x'}_{\perp})} 
\nonumber
\\
&&\times\underbrace{\int d^2{\bf k'}_{\perp}e^{i{\bf k'}_{\perp}\cdot {\bf x}_{\perp} - i\frac{{\bf k'}_{\perp}^2}{4k_-} (U-U_0) - i{\bf k'}_{\perp}\cdot {\bf x'}_{\perp}}}_{I_{{\bf k'}_{\perp}}},
\nonumber
\\
&&{\textrm{for}}~U>U_0~.
\label{2.03}
\end{eqnarray}
(Also see \cite{Klimcik:1989kh}).
Here the wave vector $k^a=(k_+,k_-,{\bf k}_{\perp})$ with $k_{\pm} = (1/2)(k_T\pm k_X)$ and ${\bf k^2}_{\perp} = k_Y^2+k_Z^2$ with $k_+=({\bf k}_{\perp}^2/4k_-)$. The normalization is given by $N_k = [(2\pi)^{3/2}\sqrt{2k_-}]^{-1}$. Note that in the above, integration over ${\bf k'}_{\perp}$; i.e. $I_{{\bf k'}_{\perp}}$ can be evaluated (See Appendix \ref{App1}). This yields
\begin{eqnarray}
&&f_{k_-, {\bf k}_{\perp}} = - N_k\frac{4i\pi k_-}{(U-U_0)}e^{-ik_-V}
\nonumber
\\
&&\times\int \frac{d^2{\bf x'}_{\perp}}{(2\pi)^2} e^{i{\bf k}_{\perp}\cdot {\bf x'}_{\perp}+ik_-f({\bf x'}_{\perp}) + \frac{ik_-|{\bf x}_{\perp}-{\bf x'}_{\perp}|^2}{U-U_0}}~.
\label{2.07}
\end{eqnarray}
Below we shall use this expression for calculating the positive frequency Wightman function. 


{\it Wightman function} --
The massless scalar field, in terms of its modes, can be decomposed as
\begin{equation}
\phi(x) = \int{dk_- d^2{\bf k}_\perp} \Big[a_{k_-,{\bf k}_\perp}f_{k_-, {\bf k}_{\perp}} + a^\dagger_{k_-,{\bf k}_\perp}f^*_{k_-, {\bf k}_{\perp}}\Big]~,
\label{2.08}
\end{equation} 
with $k_->0$ and ${\bf k}_{\perp}$ is arbitrary.
The positive frequency Wightman function $W^{(+)}(x_2;x_1) = \bra{0}\phi(x_2)\phi(x_1)\ket{0}$ by (\ref{2.08}) turns out to be 
\begin{equation}
W^{(+)}(x_2;x_1)=\int{dk_{-}d^2{\bf k}_{\perp}}f_{k_{-}, {\bf k}_{\perp}} (x_2) f^*_{k_{-}, {\bf k}_{\perp}} (x_1)~,
\label{2.12}
\end{equation}
where $\bra{0}a_{k_{2-},{\bf k}_{2\perp}} a^\dagger_{k_{1-},{\bf k}_{1\perp}}\ket{0} = \delta(k_{2-}-k_{1-})\delta^{(2)}({\bf k}_{2\perp} - {\bf k}_{1\perp})$ has been used.
Here  $T_2>T_1$ and $\ket{0}$ is SMB vacuum in presence of shock wave. 

Since the scalar modes (\ref{2.02}) for $U<U_0$ are exactly same as those for the usual Minkowski spacetime (i.e. in absence of shock wave), $W^{(+)}(x_2;x_1)$ is given by the our well known Minkowski expression \cite{Book1}:
\begin{equation}
W^{(+)}(x_2;x_1) = -\frac{1}{4\pi^2}\frac{1}{\Delta T^2 - |\Delta{\bf X}|^2}~,
\label{2.10}
\end{equation} 
where we used the notation $\Delta X^a= X^a_2 -X^a_1$. On the other hand mode function in the region $U>U_0$ is modified due to shock wave. But still computation shows that one obtains the same expression (\ref{2.10})  (See Appendix \ref{App2} for details).

Hence we observed that the positive frequency Wightman function for the massless scalar field does not modify due to the presence of shock wave of this particular form. On both sides of the shock wave location (here it is $U=U_0$), the two point function does not change. Therefore any quantity which is being calculated from two point function will not modify and retain its original value. This explains why the accelerated observer does not see any alteration of the Unruh temperature. 

In \cite{Compere:2019rof}, it has been shown by calculating Bogoliubov coefficients that Unruh temperature does not change while the relevant Bogoliubov coefficient, which related to particle number, changes by a phase factor. Therefore the particle number, as seen by the accelerated observer in SMB vacuum, does not change. Note that in their analysis the acceleration was taken along $X$ axis. So it was not clear whether there is any non-triviality that can happen if the observer is moving along any of the transverse directions. This is a very relevant question, as the shock wave gives rise to modification of the metric along the transverse directions (see metric (\ref{2.01})). Now the present analysis does have this answer. Since the Wightman function is not changing, the spectrum of particles, seen by the uniformly accelerated observer in the SMB vacuum, remains same whatever the direction of motion. Moreover, it tells that the usual results for other types of observer (like rotating) on Minkowski spacetime also hold in this case as well. Not only that the different physical quantities which are evaluated from two point function, like renormalised stress-tensor, also do not give the signature of presence of wave. 

{\it Comparison of vacua on both sides} --
Now we shall investigate how the vacuum of one side of shock wave location looks with respect to other side of it. This will be investigated by calculating the Bogoliubov coefficients which connect the massless scalar modes on both sides. Let us express $U<U_0$ modes in terms of the modes for $U>U_0$:
\begin{eqnarray}
&&g_{k_-,{\bf k}_\perp}(x) = \int dk'_- d^2{\bf k'}_\perp\Big[\alpha({k_-,{\bf k}_\perp;k'_-,{\bf k'}_\perp}) f_{k'_-,{\bf k'}_\perp}(x) 
\nonumber
\\
&&+ \beta({k_-,{\bf k}_\perp;k'_-,{\bf k'}_\perp}) f^*_{k'_-,{\bf k'}_\perp}(x)\Big]~,
\label{2.18}
\end{eqnarray}  
where the Bololiubov coefficients are found to be 
\begin{eqnarray}
&&\alpha({k_-,{\bf k}_\perp;k'_-,{\bf k'}_\perp}) =-i\int dX d^2{\bf x}_\perp~ \Big[g_{k_-,{\bf k}_\perp}\partial_T f^*_{k'_-,{\bf k'}_\perp} 
\nonumber
\\
&&- \partial_T(g_{k_-,{\bf k}_\perp})f^*_{k'_-,{\bf k'}_\perp}\Big]_{T=0} 
\nonumber
\\
&=& \frac{1}{(2\pi)^2}\delta(k_- -k'_-)\int d^2{\bf x'}_{\perp} e^{- i({\bf k'}_{\perp} - {\bf k}_{\perp}) \cdot{\bf x'}_{\perp}-ik'_-f({\bf x'}_\perp)}~.
\nonumber
\\
\label{2.26} 
\end{eqnarray}
and
\begin{eqnarray}
&&\beta({k_-,{\bf k}_\perp;k'_-,{\bf k'}_\perp}) =i\int dX d^2{\bf x}_\perp~ \Big[g_{k_-,{\bf k}_\perp}\partial_T f_{k'_-,{\bf k'}_\perp}
\nonumber
\\
&& - \partial_T(g_{k_-,{\bf k}_\perp})f_{k'_-,{\bf k'}_\perp}\Big]_{T=0} = 0~.
\label{2.27} 
\end{eqnarray}
The detail computation is given in Appendix \ref{App3}. For the consistency, with (\ref{2.26}) one can check that 
\begin{eqnarray}
&&\int dk'_- d^2{\bf k'}_\perp \alpha({k_{1-},{\bf k}_{1\perp};k'_-,{\bf k'}_\perp}) \alpha^*({k_{2-},{\bf k}_{2\perp};k'_-,{\bf k'}_\perp})
\nonumber
\\
&=& \delta(k_{2-} - k_{1-})\delta^{(2)}({\bf k}_{2\perp} - {\bf k}_{1\perp})~.
\label{2.28}
\end{eqnarray}
Now since other coefficient vanishes, the relation among the Bogoliubov coefficients \cite{Book1}
\begin{eqnarray}
&&\int dk'_- d^2{\bf k'}_\perp \Big[\alpha({k_{1-},{\bf k}_{1\perp};k'_-,{\bf k'}_\perp}) \alpha^*({k_{2-},{\bf k}_{2\perp};k'_-,{\bf k'}_\perp})
\nonumber
\\
&& - \beta({k_{1-},{\bf k}_{1\perp};k'_-,{\bf k'}_\perp}) \beta^*({k_{2-},{\bf k}_{2\perp};k'_-,{\bf k'}_\perp}) \Big]
\nonumber
\\
&=&\delta(k_{2-} - k_{1-})\delta^{(2)}({\bf k}_{2\perp} - {\bf k}_{1\perp})~,
\end{eqnarray}
is satisfied. This verifies the consistency of our results (\ref{2.26}) and (\ref{2.27}). 

 Therefore nothing special is measured by static observer, stationed on one side of shock wave location, in the vacuum of other side.  
This implies that the vacuum states on both sides of the shock wave location are equivalent. Hence the shock hair of this type does not keep any imprint at the quantum level in this particular quantity as well.  

{\it Role of non-vacuum state in retrieving memory} --
So far we talked about the vacuum states of the scalar field, i.e. SMB vacuum and we found that this does not capture the signature of classical shock wave in the spacetime. Now we shall concentrate on the non-vacuum states of the field, say $\ket{\Psi}$. The two point function for this state is given by \cite{Lochan:2014xja}
\begin{eqnarray}
&&C(x_2;x_1)=\bra{\Psi}\phi(x_2)\phi(x_1)\ket{\Psi}
\nonumber
\\
&=& W^{(+)}(x_2;x_1) + 2\textrm{Re} \Big(\Phi_{\textrm{eff}}(x_2)\Phi^*_{\textrm{eff}}(x_1)\Big)~,
\label{3.01}
\end{eqnarray} 
where 
\begin{equation}
\Phi_{\textrm{eff}}(x) = \int\frac{dk_- d^2{\bf k}_\perp}{(2\pi)^{3/2} \sqrt{2k_-}}h(k_-,{\bf k}_\perp) f_{k_-,{\bf k}_\perp}(x)~.
\label{3.02}
\end{equation}
In the above, $h(k_-,{\bf k}_\perp)$ is related to probability distribution of scalar field momentum. So this consists of two parts. One part is due to vacuum expectation value and other is due to non-vacuum effect. We already discussed that the vacuum part does not attribute the presence of shock wave in the spacetime. Now we shall investigate whether the non-vacuum contribution can capture the shock wave effect.

For simplicity, here we consider a one particle state of definite momentum $k^0$ in Minkowski spacetime. Then $h(k_-,{\bf k}_\perp)$ is given by \cite{Lochan:2014xja}
\begin{equation}
h(k_-,{\bf k}_\perp)= (2\pi)^{3} \sqrt{2k_-}~\delta(k_- - k_-^0)\delta^{(2)}({\bf k}_\perp - {\bf k^0}_\perp)~,
\label{3.03}
\end{equation}
and with this (\ref{3.02}) turns out to be
\begin{equation}
\Phi_{\textrm{eff}}(x) = (2\pi)^{3/2}f_{k_-^0,{\bf k^0}_\perp}(x)~.
\label{3.04}
\end{equation}
Therefore using (\ref{2.07}) one finds
\begin{eqnarray}
\Phi_{\textrm{eff}}(x_2)\Phi^*_{\textrm{eff}}(x_1)&=&|N_{k^0}|^2\frac{(2\pi)^316\pi^2 (k_-^0)^2}{(U_2-U_0)(U_1-U_0)}
\nonumber
\\
&\times& e^{-ik_-^0(V_2-V_1)} I_{\bf x_2}I^*_{\bf x_1}~,
\label{3.05}
\end{eqnarray}
where
\begin{equation}
I_{\bf x_n} = \int\frac{d^2{\bf x'}_\perp}{(2\pi)^2}e^{i{\bf k^0}_\perp\cdot{\bf x'}_\perp + ik_-^0 f({\bf x'}_\perp) + ik_-^0\frac{|{\bf x}_{n\perp} - {\bf x'}_\perp|^2}{U_n-U_0}}~.
\label{3.06}
\end{equation}	
Here we have $n\in 1,2$. Although the above integration can not be done without the explicit form of shock wave contribution $f({\bf x}_\perp)$, it is obvious that it picks the influence of wave. Therefore any observable quantity, calculated from this two point correlator for one particle state, can reflect the classical memory of the spacetime.

Now in order to investigate whether shock wave does give any impression on the two point function (\ref{3.01}), we need to evaluate integration (\ref{3.06}). Just to have a feel of this here we choose a very particular type of shock wave for which $f({\bf x}_\perp)$ is known. For simplicity we consider the following form of the function \cite{Klimcik:1988az} 
\begin{equation}
f({\bf x}_\perp) = -a(Y^2+Z^2)~,
\label{3.07}
\end{equation}
where $a$ is a constant and $Y, Z$ are transverse coordinates. This represents an infinite planar shell of null matter with constant energy density. Using this in (\ref{3.06}) and then performing the integration we obtain (see Appendix \ref{App4} for details of the derivation)  
\begin{eqnarray}
&&\textrm{Re} \Big(\Phi_{\textrm{eff}}(x_2)\Phi^*_{\textrm{eff}}(x_1)\Big) = \frac{(2\pi)^3|N_{k^0}|^2}{(U_2-U_0)(U_1-U_0)}
\nonumber
\\
&\times&\frac{1}{(a-\frac{1}{U_2-U_0})(a-\frac{1}{U_1-U_0})}\cos(A_2-A_1)~,
\label{3.09}
\end{eqnarray}
where
\begin{eqnarray}
A_n=-k_-^0V_n + \frac{{\bf k^0}_\perp^2 - \frac{4k_-^0}{U_n-U_0}{\bf k^0}_\perp\cdot{\bf x}_{n\perp} + \frac{4a(k_-^0)^2}{U_n-U_0}{\bf x}_{n\perp}^2}{4k_-^0(a-\frac{1}{U_n-U_0})}~.
\label{3.12}
\end{eqnarray}
This result is very interesting as it captures the effects of shock wave in the spacetime. One can check that for $a\rightarrow 0$ limit, the above reduces to the usual result (i.e. in absence of shock wave), given in \cite{Lochan:2014xja} which is valid in the region $U<U_0$ here. 
An interesting point to be noted is that (\ref{3.12}) in $U>U_0$ is completely distinct from that in $U<U_0$. Therefore although the vacuum is equivalent on both sides, however the non-vacuum states are not.	
So any observable, calculated from (\ref{3.12}) must give the signature of this non-triviality. For example, the response function \cite{State1} of a static Unruh - De-Witt detector for $U<U_0$ is given by Dirac-delta function $\delta(\Delta E - \omega^0)$, where $\Delta E$ is the gap between the detector's energy levels and $\omega^0 = k_+^0+k_-^0=k_T^0$. So it will measure a spectrum which is sharply picked at $\omega^0$. In this case the two point function is time translation invariant and so the system is in equilibrium. On the other hand, response function in region $U>U_0$ will have a distribution which is not Dirac-delta function as $a\neq 0$ and also it depends on time. In this case two point function is not time translation invariant and hence the system is not in equilibrium. Numerically, one can check that the peak of the distribution decreases with the increase of $a$ (see Appendix \ref{App5}). This implies that the response function indeed captures the presence of classical shock wave.
Another aspect of the above result is, the observables are dependent on movement of direction of the frame -- the transverse direction and the $X$ direction give different behaviour when $a\neq 0$. This anisotropy is absent for $a=0$ case.

{\it Conclusions} --
In this letter, we studied the quantum consequences of presence of localised shock wave in the Minkowski spacetime. A particular type of metric, given by Dray and 't Hooft, has been considered here. The SMB can be regarded as the modified version of Minkowski spacetime due to the back reaction of wave propagating on this. With the idea that various physical quantities can be extracted from two point function for fields, we computed the two point function of massless scalar field for this wave modified Minkowski metric. So if there is any signature of wave in the correlator, that might be observable in the physical quantities. 

The computation showed that the vacuum correlator does not capture the classical shock wave soft hair, thereby any observer in this SMB vacuum will not feel the wave at the quantum regime. Moreover, we found that the vacuum structure on both sides of wave is equivalent.  

Finally, we observed that the non-vacuum contribution in the non-vacuum two point function for the future side of the wave is modified due to propagation of shock wave in the spacetime; whereas on the other side it is trivial. Here for simplicity we did computations in one particle state. This implies that the one particle state on both sides are inequivalent. Due to this non-triviality, we conclude that although vacuum can not give the signature of classical memory of spacetime, the non-vacuum states can provide such memory. Therefore, it may be the case that the shock wave not only changes the background, but also causes the change of the system so that the fields are no longer in vacuum state; rather they are in exited states. So the classical memory is now encoded in non-vacuum states. This idea is similar to what was taken in \cite{Lochan:2015oba} to enlighten the information extraction of the initial state.
Therefore we conjecture that {\it non-vacuum states of fields are more relevant for retrieval of the memory}.
\vskip 5mm
{\it Acknowledgments}: {\sc This work is dedicated to those who are helping us to fight against COVID-19 across the globe.}

The author thanks Krishnakanta, Surojit and Sumit for helpful suggestions and comments on the first draft of the manuscript.

\vskip 3mm

\begin{widetext}

\begin{center}
{\underline{\bf Supplementary material}}
\end{center}
\appendix
\section{Evaluation of Eq. (\ref{2.07})}\label{App1}
$I_{{\bf k'}_{\perp}}$ in (\ref{2.03}) can be expressed as
\begin{equation}
I_{{\bf k'}_{\perp}}= \int_{-\infty}^{+\infty} dk'_Y e^{ik'_Y(Y-Y') - i\frac{U-U_0}{4k_-}k'^2_{Y}}\int_{-\infty}^{+\infty} dk'_Z e^{ik'_Z(Z-Z') - i\frac{U-U_0}{4k_-}k'^2_Z}~.
\label{2.04}
\end{equation}
Each integration is identical and can be done by the following general result
\begin{equation}
\int_{-\infty}^{+\infty}d\sigma e^{a\sigma-b\sigma^2} = \sqrt{\frac{\pi}{b}}~e^{\frac{a^2}{4b}}~.
\label{2.05}
\end{equation}
Then we find
\begin{equation}
I_{{\bf k'}_{\perp}} = -\frac{4i\pi k_-}{(U-U_0)}e^{\frac{ik_-|{\bf x}_{\perp}-{\bf x'}_{\perp}|^2}{U-U_0}}~.
\label{2.06}
\end{equation}
With this the massless scalar mode (\ref{2.03}) for $U>U_0$ becomes (\ref{2.07}).

\section{Evaluation of Eq. (\ref{2.17})}\label{App2}
Substituting (\ref{2.07}) in (\ref{2.12}) and rearranging we obtain
\begin{eqnarray}
W^{(+)}(x_2;x_1) &=& \frac{1}{\pi (2\pi)^4 (U_2-U_0)(U_1-U_0)}\int dk_-k_-\int d^2{\bf x'}_{2\perp}d^2{\bf x'}_{1\perp}e^{-ik_-(V_2-V_1) + ik_-\Big(f(\bf x'_{2\perp}) - f(\bf x'_{1\perp})\Big)}
\nonumber
\\
&\times& e^{ik_-\Big(\frac{|{\bf x}_{2\perp}- {\bf x'}_{2\perp}|^2}{U_2-U_0} - \frac{|{\bf x}_{1\perp}- {\bf x'}_{1\perp}|^2}{U_1-U_0} \Big)} 
\underbrace{\int d^2{\bf k}_{\perp} e^{i{\bf k}_\perp\cdot({\bf x'}_{2\perp}- {\bf x'}_{1\perp})}}_{(2\pi)^2\delta^{(2)}({\bf x'}_{2\perp}- {\bf x'}_{1\perp})}
\nonumber
\\
&=& \frac{4\pi}{(2\pi)^4 (U_2-U_0)(U_1-U_0)}
\int dk_-k_-e^{-ik_-(V_2-V_1)}
\underbrace{\int d^2{\bf x'}_{\perp} e^{ik_-\Big(\frac{|{\bf x}_{2\perp}- {\bf x'}_{\perp}|^2}{U_2-U_0} - \frac{|{\bf x}_{1\perp}- {\bf x'}_{\perp}|^2}{U_1-U_0} \Big)}}_{I_{{\bf x'}_\perp}}~.
\label{2.13}
\end{eqnarray}
$I_{{\bf x'}_\perp}$ can be evaluated in the following way. This is decomposed into two identical integration:
\begin{equation}
I_{{\bf x'}_\perp} = \int_{-\infty}^{+\infty}dY' e^{ik_-\Big(\frac{(Y_2-Y')^2}{U_2-U_0} -
	\frac{(Y_1-Y')^2}{U_1-U_0}\Big)}
\int_{-\infty}^{+\infty}dZ' e^{ik_-\Big(\frac{(Z_2-Z')^2}{U_2-U_0} - \frac{(Z_1-Z')^2}{U_1-U_0}\Big)}~.
\label{2.14}
\end{equation}
Using a general result
\begin{equation}
\int_{-\infty}^{+\infty}d\sigma e^{a(b-\sigma)^2 - c(d-\sigma)^2} = \sqrt{\frac{\pi}{c-a}}~e^{-\frac{ac}{a-c}(b-d)^2}~,
\label{2.15}
\end{equation}
one obtains
\begin{equation}
I_{{\bf x'}_\perp} = \frac{i\pi}{k_-\Big(\frac{1}{U_2-U_0} - \frac{1}{U_1-U_0}\Big)}\exp\Big[\frac{ik_-|{\bf x}_{2\perp} - {\bf x}_{1\perp}|^2}{(U_2-U_1)} \Big]~.
\end{equation}
Substitution of this in (\ref{2.13}) yields 
\begin{equation}
W^{(+)}(x_2;x_1) = -\frac{4i\pi^2}{ (2\pi)^4 (U_2-U_1)}
\int_{0}^{\infty} dk_-
\exp\Big[-ik_-\Big((V_2-V_1)-\frac{|{\bf x}_{2\perp} - {\bf x}_{1\perp}|^2}{U_2-U_1}\Big)\Big]~,
\label{2.17n}
\end{equation}
which after $k_-$ integration gives rise to 
\begin{equation}
W^{(+)}(x_2;x_1) = -\frac{1}{4\pi^2}\frac{1}{(\Delta U)(\Delta V)-\Delta Y^2 -\Delta Z^2}~.
\label{2.17}
\end{equation}
Next using $U=T-X$, $V=T+X$, the above reduces to the standard form (\ref{2.10}).

\section{Evaluation of Eq. (\ref{2.26})}\label{App3}
Using (\ref{2.03}) one can calculate the following quantity:
\begin{equation}
\partial_T f^*_{k'_-,{\bf k'}_\perp}|_{T=0} = N_{k'}^* e^{ik'_-X} \int \frac{d^2{\bf x'}_{\perp}}{(2\pi^2)} e^{-i{\bf k'}_{\perp}\cdot{\bf x'}_\perp - i k'_-f({\bf x'}_\perp)}\int d^2{\bf k''}_\perp e^{-i{\bf k''}_{\perp}\cdot{\bf x}_\perp - \frac{i{\bf k''^2}_{\perp}}{4k'_-}(X+U_0) + i{\bf k''}_{\perp}\cdot{\bf x'}_\perp} (ik'_- + \frac{i{\bf k''}_\perp^2}{4k'_-}).
\label{2.21}
\end{equation}
Then after rearrangement one finds
\begin{eqnarray}
&&\int dX d^2{\bf x}_\perp~ g_{k_-,{\bf k}_\perp}\partial_T f^*_{k'_-,{\bf k'}_\perp}\Big|_{T=0} = N_k N_{k'}^*\int \frac{d^2{\bf x'}_{\perp}d^2{\bf k''}_{\perp}}{(2\pi)^2}e^{ik_+U_0 - i{\bf k'}_{\perp}\cdot{\bf x'}_{\perp}-ik'_-f({\bf x'}_\perp)-\frac{i{\bf k''^2}_{\perp}}{4k'_-}U_0 + i{\bf k''}_{\perp}\cdot{\bf x'}_{\perp}}
\nonumber
\\ 
&\times&(ik'_- + \frac{i{\bf k''}_\perp^2}{4k'_-})
\underbrace{\int dX e^{-ik_- X + ik'_-X + ik_+X - \frac{i{\bf k''^2}_{\perp}}{4k'_-}X}}_{2\pi \delta\Big(-k_-+k'_-+k_+-\frac{{\bf k''}_\perp^2}{4k'_-}\Big)}
\underbrace{\int d^2{\bf x}_\perp e^{i{\bf k}_{\perp}\cdot{\bf x}_\perp- i{\bf k''}_{\perp}\cdot{\bf x}_\perp}}_{(2\pi)^2 \delta^{(2)}({\bf k}_\perp - {\bf k''}_\perp)}
\nonumber
\\
&=& (2\pi) N_k N_{k'}^* e^{i\Big(k_+ - \frac{{\bf k}_\perp^2}{4k'_-}\Big)U_0}\Big(ik'_- + \frac{i{\bf k}_\perp^2}{4k'_-}\Big)\delta\Big(k_--k'_--k_++\frac{{\bf k}_\perp^2}{4k'_-}\Big)
\int d^2{\bf x'}_{\perp} e^{- i{\bf k'}_{\perp}\cdot{\bf x'}_{\perp}-ik'_-f({\bf x'}_\perp) + i{\bf k}_{\perp}\cdot{\bf x'}_{\perp}}~.
\label{2.22}
\end{eqnarray}
Next using the fact that
\begin{equation}
\delta\Big(k_--k'_--k_++\frac{{\bf k}_\perp^2}{4k'_-}\Big) = \frac{\delta(k'_- - k_-)}{1+\frac{k_+}{k_-}} + \frac{\delta(k'_- + k_+)}{1+\frac{k_-}{k_+}}~,
\label{2.23}
\end{equation}
where one introduces ${\bf k}_\perp^2 = 4k_+k_-$, with $k_-,k_+>0$, we obtain
\begin{equation}
\int dX d^2{\bf x}_\perp~ g_{k_-,{\bf k}_\perp}\partial_T f^*_{k'_-,{\bf k'}_\perp}\Big|_{T=0} = (2\pi)|N_k|^2 (ik_-)\delta(k_- -k'_-)\int d^2{\bf x'}_{\perp} e^{- i({\bf k'}_{\perp} - {\bf k}_{\perp}) \cdot{\bf x'}_{\perp}-ik'_-f({\bf x'}_\perp)}~.
\label{2.24} 
\end{equation}
In the above, as $k_-,k_+>0$, only first part of (\ref{2.23}) contributes.
Similarly we find
\begin{equation}
\int dX d^2{\bf x}_\perp~ \partial_T(g_{k_-,{\bf k}_\perp})f^*_{k'_-,{\bf k'}_\perp}\Big|_{T=0} = (2\pi)|N_k|^2 (-ik_-)\delta(k_- -k'_-)\int d^2{\bf x'}_{\perp} e^{- i({\bf k'}_{\perp} - {\bf k}_{\perp}) \cdot{\bf x'}_{\perp}-ik'_-f({\bf x'}_\perp)}~.
\label{2.25} 
\end{equation}
Substitution of (\ref{2.24}) and (\ref{2.25}) in the first equality of (\ref{2.26}) and normalization $N_k = \frac{1}{(2\pi)^{3/2}\sqrt{2k_-}}$ gives the final form. Likewise Eq. (\ref{2.27}) can also be derived.

\section{Derivation of Eq. (\ref{3.09})}\label{App4}
Using (\ref{3.07}) in (\ref{3.06}) and rearranging we write in Cartesian form as
\begin{equation}
I_{\bf x_n} = \frac{1}{(2\pi)^2}\int_{-\infty}^{+\infty} dY' e^{ik_Y^0Y'-iak_-^0Y'^2 + \frac{ik_-^0}{U_n-U_0}(Y-Y')^2}
\int_{-\infty}^{+\infty} dZ' e^{ik_Z^0Z'-iak_-^0Z'^2 + \frac{ik_-^0}{U_n-U_0}(Z-Z')^2}~.
\label{3.10}
\end{equation}
In the above both the integrals are identical. They can be evaluated by the following general formula
\begin{equation}
\int_{-\infty}^{+\infty}d\sigma e^{A\sigma-B\sigma^2+C(D-\sigma)^2} = \sqrt{\frac{\pi}{B-C}}~e^{\frac{A^2-4ACD+4BCD^2}{4(B-C)}}~.
\label{3.11}
\end{equation}
This yields
\begin{equation}
I_{\bf x_n} = \frac{1}{4i\pi k_-^0 (a-\frac{1}{U_n-U_0})} 
\exp\Big[ \frac{i{\bf k^0}_\perp^2 - \frac{4ik_-^0}{U_n-U_0}{\bf k^0}_\perp\cdot{\bf x_n}_\perp + \frac{4iak_-^{0^2}}{U_n-U_0}{\bf x_n}_\perp^2}{4k_-^0(a-\frac{1}{U_n-U_0})}\Big]~.
\label{3.08}
\end{equation}  
Substituting the above in (\ref{3.05}) we find (\ref{3.09}).

\section{Response function}\label{App5}
The response function is defined as \cite{Lochan:2014xja}
\begin{equation}
\frac{dR(\Delta E)}{d\tau'} \sim \int_{-\infty}^{+\infty} d\tau e^{-i \Delta E \tau}\bra{\Psi}\phi(x_2)\phi(x_1)\ket{\Psi}~,
\label{R1}
\end{equation}
where $\tau=1/2(\tau_2-\tau_1)$, $\tau' = 1/2(\tau_2+\tau_1)$. Here $\tau$ denotes the proper time of the detector.
For static observer $\tau=T$ and the vacuum part of the correlation function (\ref{3.01}) does not contribute. So we will concentrate only on the non-vacuum part (\ref{3.09}). For the static observer (\ref{3.09}) turns out to be
\begin{eqnarray}
\textrm{Re}(\Phi(x_2)\Phi^*(x_1)) = \frac{1}{2k_-^0\Big[a(\tau+T')-1\Big] \Big[a(-\tau+T')-1\Big]} \cos\Big[-2k_-^0 \tau + k_+^0\Big(\frac{1}{a-\frac{1}{\tau+T'}} - \frac{1}{a-\frac{1}{-\tau+T'}} \Big) \Big]~,
\label{R2}
\end{eqnarray}
where $T' = \tau'-U_0$
Substitution of this in (\ref{R1}) yields our desire quantity. We numerically integrate this and plot below $dR/d\tau'$ VS $\Delta E$ for different values of $a$.
\begin{figure}[!ht]
	\centering
	\includegraphics[scale=0.50, angle=0]{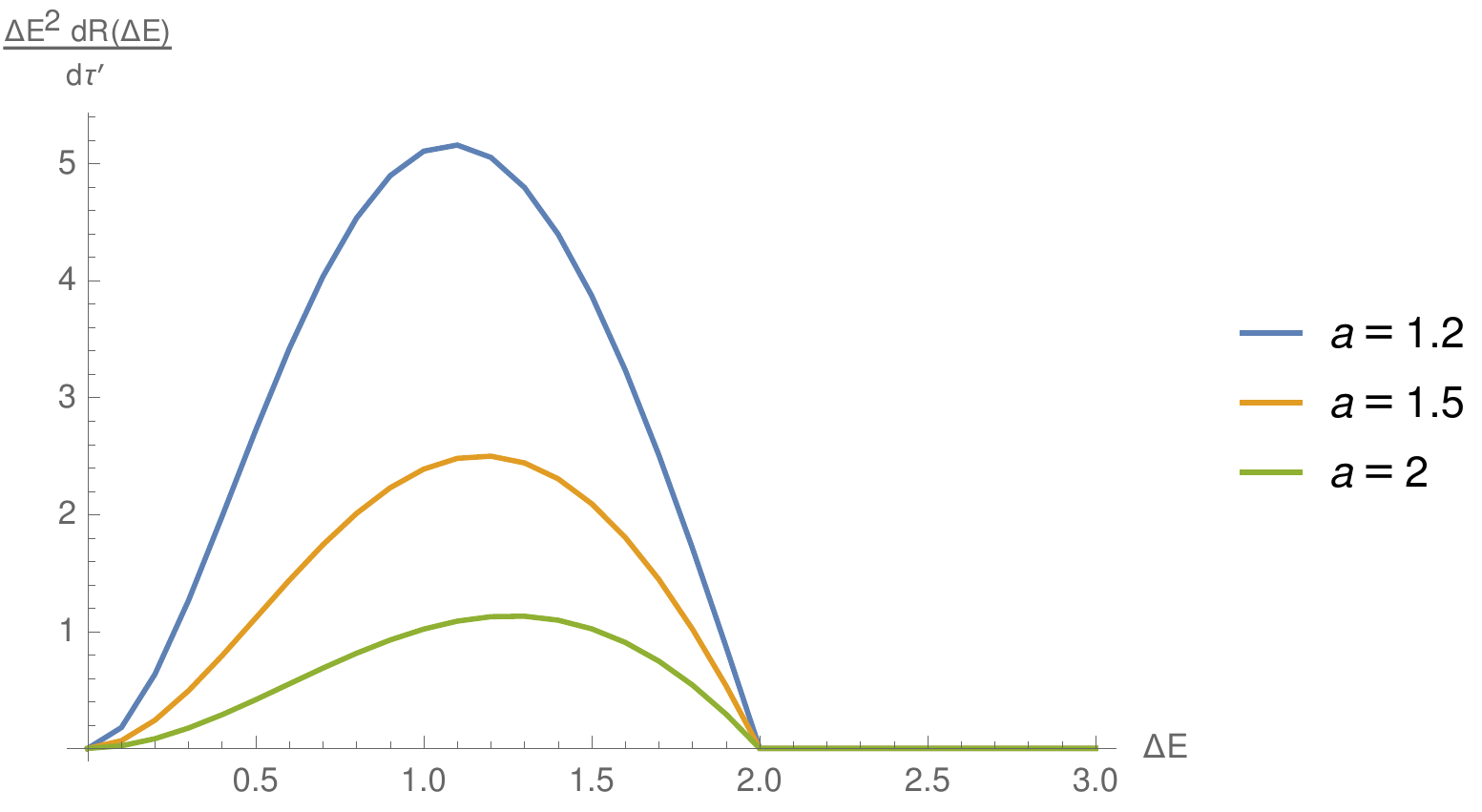}
	\caption{(Color online) $dR/d\tau'$ VS $\Delta E$ plot for different values of $a$. The other parameters are kept constant to values $k_-^0 =1, k_+^0 =1.5$ and $T'=1$. For numerical purpose we replaced $\tau$ by $\tau-i\epsilon$ with $\epsilon = 0.05$ to make the integration a convergent one.}
	\label{Fig1}
\end{figure}
This shows that the peak of the response function decreases with the increase of strength of the shock wave, represented by $a$.
\end{widetext}

\end{document}